\documentstyle[12pt]{article}
\newtheorem{theorem}{Theorem}[section]

\newtheorem{remark}[theorem]{Remark}   

\newfont{\BB}{msbm10}
\def\R{\mbox{\BB R}}

\begin{document}
\title{A footnote to Nelson's interpretation of the two-slit experiment\\
     {\small{\em
dedicated  to {\em Francesco Guerra} on the occasion of his 60th
birthday}}}
\author{Michele Pavon\\Dipartimento di Matematica Pura e Applicata\\
Universit\`a di Padova\\ via Belzoni 7, and ISIB-CNR\\ 35131 Padova,
Italy\\{\tt pavon@math.unipd.it}}

\maketitle
\begin{abstract} We seek to complement Nelson's work on the two-slit
experiment by showing that the {\em two-slit process}, whose density exhibits
the characteristic interference pattern, may be obtained as the model after the
beam has reached the screen by means of a variational mechanism.  The {\em
one-slit process}, modeling the beam before it reaches the screen, plays the
role of a {\em reference model}.
\end{abstract}

\section {Introduction} When a beam of electrons originating from a
source ${\cal S}$ goes through a crystal, one observes a diffraction
pattern resembling that characteristic of interference in wave motion. In
a celebrated {\em gedanken experiment} \cite[pp.2-5]{FH}, the crystal is
replaced by a screen with two slits. As it is well-known, according to
quantum mechanics, this is an instance of a {\em superposition} of two
states. The two possibilities of going through the top or the bottom
slit interfere. Hence, the probability of arrival at the second screen is
not simply the sum of the probabilities of coming through the top and the
bottom slit.

In
\cite{N2,N3}, Edward Nelson made a remarkable calculation showing  that,
according to his version of stochastic mechanics, the
observed interference effect
     can be explained within the frame of classical
probability. In particular, the probability of arrival
is indeed the sum of the probabilities of coming through the top and
the bottom slit.

In this note we seek to complement what Nelson did. We model the
emission from the source through a suitable Gaussian process ({\em one-slit
process}).  We then take the latter as a {\em reference} process in a
variational
problem that takes into account the presence of the screen with the two
slits. The solution of the variational problem, suitably extending on the
results of \cite{P1,P2}, is the stochastic process  that Nelson has
analyzed ({\em two-slit process}) whose density profile  is familiar in wave
interference.

\section{Elements of the Nelson-Guerra stochastic mechanics}
Nelson's stochastic mechanics is a quantization procedure for classical
dynamical systems based on diffusion processes. Following some early
work by Feynes
\cite{Fe} and others, Nelson and Guerra elaborated a clean formulation
starting from 1966
\cite{N0,N1,G} and references therein, showing that the Schr\"{o}dinger
equation could be derived from a continuity type equation plus a Newton type
law, provided one accepted a certain  definition for the stochastic
acceleration.
The Newton-Nelson law was later shown to follow, in analogy to classical
mechanics, from a Hamilton-like stochastic variational principle \cite{Y,GM}.
Other versions of the variational principle have been proposed in
\cite{N2,BCZ,P0,ROS}.

\noindent
Consider the case of a
nonrelativistic particle of mass $m$. Let
$\{\psi(x,t); t_0\le t\le t_1\}$, satisfying  the {\it Schr\"{o}dinger
equation}
\begin{equation}\label{H7} \frac{\partial{\psi}}{\partial{t}} =
\frac{i\hbar}{2m}\Delta\psi -
\frac{i}{\hbar}V(x)\psi, \end{equation}
be such that
\begin{equation}\label{FA}||\nabla\psi||^2_2\in
L^1_{{\rm loc}}[t_0,+\infty).
\end{equation}
This is Carlen's
{\em finite action condition}. Under these hypotheses, the Nelson measure
$P$
may be constructed on path space, \cite{C},\cite{Car}, \cite [Chapter
IV]{BCZ}, and references therein.
Namely, letting  $\Omega:={\cal C}([t_0,t_1],\R^n)$ the $n$-dimensional
continuous functions on $[t_0,t_1]$, under the  probability measure $P$,
the canonical coordinate process $x(t,\omega)=\omega(t)$ is an
$n$-dimensional, Markov,
finite-energy diffusion process $\{x(t);t_0\le t\le t_1\}$,
called {\em Nelson's process}, having
(forward) Ito differential
\begin{equation}\label{N}
dx(t)=\left[\frac{\hbar}{m}\nabla\left(\Re
\log\psi(x(t),t) + \Im \log\psi(x(t),t)\right)\right]dt
+\sqrt{\frac{\hbar}{m}}dw(t),
\end{equation}
where $w$ is a standard, $n$-dimensional Wiener process. Moreover, the
probability density $\rho(\cdot,t)$ of
$x(t)$ satisfies Born's relation
\begin{equation}\label{D}\rho(x,t)=|\psi(x,t)|^2,\quad \forall t \in [t_0,t_1].
\end{equation}

\section{Nelson's treatment of the two-slit experiment}
We recall some essential points in Nelson's analysis in
\cite{N2,N3}. Let $x\in\R^3$, and consider {\em free} motions, i.e.
$V\equiv 0$, in a frame of reference comoving with the beam. For the sake
of simplicity, take
$m=1$,
$\hbar=1$ and $t_0=0$.   The slits are located at $\pm a$, where
$a$ is a vector in $\R^3$.  Let
$$\psi_0(x)=\left(\frac{\lambda}{\pi}\right)^{1/4}\exp(-\frac{|x|^2}{2
\lambda})$$
be the initial condition at time $t_0$. Then, the corresponding solution
of
\begin{equation}\label{0}
\frac{\partial{\psi}}{\partial{t}} =
\frac{i\hbar}{2m}\Delta\psi ,
\end{equation}
is
\begin{equation}\label{TS2}\psi_0(x,t)=\left(\frac{\lambda}{\pi}\right
)^{1/4}(\lambda+it)^{-1/2}\exp
\left(-\frac{|x|^2}{2(\lambda+it)}\right).
\end{equation}
Nelson calls the {\em one-slit process} the Gaussian process associated to
this evolution whose forward and backward drifts are given by
$$b_+(x,t)=\frac{t-\lambda}{\lambda^2+t^2}x,\quad
b_-(x,t)=\frac{t+\lambda}{\lambda^2+t^2}x.
$$
He then takes as initial condition for (\ref{H7}) the vector
\begin{equation}\label{TS1}\psi_1(x)=\gamma\left(\psi_0(x-a)+\psi_0(x+
a)\right),
\end{equation}
where $\gamma$ is a constant (close to $1/\sqrt{2}$ if $|a|$ is large
with respect to $\lambda$) to make $\psi_1$ a unit vector in $L^2$. By the
linearity of the Schr\"{o}dinger equation, the corresponding solution is
\begin{equation}\label{2s}\psi_1(x,t)=\gamma\left(\psi_0(x-a,t)+\psi_0
(x+a,t)\right).
\end{equation}
Nelson goes ahead and computes the forward drift of the {\em
two-slit} process corresponding to $\psi_1$. As only the direction
joining the slits is of interest, the problem can be
reduced to a one-dimensional problem (hence, from now on, $a,x\in\R$). He
finds that, for small times, the drift is nearly the same as for the one-slit
process. For larger times, however, the drift of the two-slit process becomes
enormous in certain regions, repelling particles from there. As a result, the
probability density of the two-slit process after some time exhibits
alternate peaks and valleys resembling those observed in wave
interference.
\section{A stochastic variational principle}
Let $\bar{v}$ be
the velocity of the beam, let $d$ be the distance between the source
and the screen with the two slits. Let $T=d/\bar{v}$ be the time it takes the
beam to reach the screen. In order to keep $t=0$ the time when the beam
is at the screen, we assume that the beam is emitted from the
source ${\cal S}$ at time $-T$. The one-slit process with $\lambda$ very
small associated to $\psi_0(x,T+t)$ as in (\ref{TS2}) models effectively
the beam of particles between times $-T$ and $0$. Suppose now we impose at
time $t=0$ the probability density
$$\rho_0(x)=\frac{1}{2}\left(|\psi_0(x-a)|^2+|\psi_0(x+a)|^2\right).
$$
We define
${\cal X}_0$
to be the
family of real-valued stochastic processes $\{x(t);-T\le t\le
0\}$ with continuous
paths and satisfying the following properties:

\begin{enumerate}
\item $x(t)$ has a nowhere vanishing probability density
$\rho(x,t)$ of class
$C^1$ for all $t\in [-T,0]$ with marginal probability density $\rho_0$ at
time $0$;
\item
$E\{\int_{-T}^{0}|\frac{\partial}{\partial
x}\log\rho(x(t),t)|^2\,dt\}<\infty$;
\item there exists for each $t\in [-T,0]$ a random variable
$v(t)$,  called the {\it
current drift}, such that, if $f(x,t)$ is of class $C^{2,1}$ satisfying
$$E\{\int_{-T}^{0}|\frac{\partial}{\partial
x} f(x(t),t)|^2\,dt\}<\infty,$$
we have
$$\frac{d E\{f(x(t),t)\}}{dt}=
E\left\{\left(\frac{\partial}{\partial t}+v(t)\frac{\partial}{\partial
x}\right)f(x(t),t)\right\}.
$$
\end{enumerate}
Notice
that property (iii)  is satisfied by a rather large class of processes
including differentiable processes, finite-energy processes with constant
diffusion coefficient \cite{F,N2}. It also includes  Markovian diffusion
processes with local diffusion coefficient given the results in
\cite{N00,Na,Mo,HP}. The  current drift $v$ is just the semi-sum of the
forward and backward drifts. Let $\cal V$ denote the family of
finite-energy stochastic processes  on $[-T,0]$. In the following
variational problem, we take one-slit process $\{x_r(t+T);-T\le t\le 0\}$
with small $\lambda$ associated to $\{\psi_0(x,T+t);-T\le t\le 0\}$ as
{\em reference} model on the time
interval $[-T,0]$. Write
$\psi_0(x,T+t)=\rho_r(x,t)\exp [\frac{i}{\hbar}S_r(x,t)]$. Rather 
than setting up a
variational problem with complex-valued drifts of the form $v-iu$ as in
Section VIII in \cite{P1} and \cite{P2}, we  formulate here a problem with
distinct real-valued velocities $v$ and $u$ as in Section IV in \cite{P1}
and
\cite{P3}. For
$(x,v,u')\in ({\cal X}_0,{\cal V},{\cal V})$, define the functional
\begin{eqnarray}\nonumber I(x,v,u):=
E\left\{\int_{-T}^{0}\left[\frac{1}{2}m\left(v(t)-
\frac{1}{m}\frac{\partial}{\partial
x} S_r(x(t),t)\right)^2\right.
\right.\\\left.\left.-
\frac{\hbar^2}{8m}\left(u'(t)-\frac{\partial}{\partial
x}\log\rho_r(x(t),t)\right)^2\right]dt-S_r(x(0),0)\right\}.
\nonumber
\end{eqnarray}
Notice that the integrand is the same as in the Guerra-Morato action
\cite{GM} where differences of  drifts replace drifts. We consider
the stochastic
differential game

$$
\min_{x\in {\cal X}_0}\max_{v\in {\cal V}}\min_{u'\in {\cal
V}}I(x,v,u')$$

subject to
\begin{eqnarray}\nonumber		v(\cdot) \;{\rm is \;the
\;current\; drift\; of}
\;x,\\u'(t)=\frac{\partial}{\partial
x}\log\rho(x(t),t)\;\forall t\;,\nonumber
\end{eqnarray}
where $\rho(\cdot,t)$ is the probability density of $x(t)$. We say
that $(x^*,v^*,u'^*)$ is a {\it saddle-point equilibrium solution} of
the game if for
all
$(x_1,v^*,u')$ and
$(x_2,v,u'^*)$ in ${\cal X}_0\times {\cal V}\times {\cal V}$
satisfying the
constraints we have
$$I(x_1,v^*,u')\le I(x^*,v^*,u'^*)\le I(x_2,v,u'^*).
$$
Let
$F:(\R\times [t_0,t_1])\rightarrow \R$ be of class $C^{1}$
satisfying the  condition in property iii). Let
$\lambda:(\R\times [t_0,t_1])\rightarrow \R$ be of class $C^1$.  For
such a pair
$(F,\lambda)$, define
    \begin{eqnarray}	\nonumber &&\Lambda^{F,\lambda}(x,v,u') =E\left\{
F(x(0),0)-F(x(-T),-T)\right.\\&&\left. -
\int_{-T}^{0}\left\{\left[\frac{\partial{F}}{\partial t} + v
\frac{\partial}{\partial x}
F\right](x(t),t)\nonumber-\lambda(x(t),t
)\left[
u'(t)-\frac{\partial}{\partial
x}\log\rho(x(t),t)\right]\right\}dt\right\}.\nonumber
\end{eqnarray}
Observe that for all triples $(x,v,u')$ in ${\cal X}_0\times
{\cal V}\times {\cal V}$
satisfying the constraints, we have $\Lambda^{F,\lambda}(x,v,u')=0$.
Obviously, if $(x,v,u')\in ({\cal X}_0\times {\cal
V}\times {\cal V})$ satisfying the constraints is a saddle-point solution for
$(I +\Lambda ^{F,\lambda})$, then it also solves the original problem with
cost function $I$.

It is possible to rewrite
$\Lambda^{F,\lambda}$ in a form more suited for our purposes using
argument similar to
that in \cite{GM}. Observe that, for any
$x\in {\cal X}_0$, we have
\begin{eqnarray}\nonumber&&
E\left\{\lambda(x(t),t)\frac{\partial}{\partial
x}\log\rho(x(t),t)\right\}
=\int_{\R^n}\lambda(x,t)\frac{\partial}{\partial
x}\rho(x,t)dx
\\&&=-\int_{\R^n}\frac{\partial}{\partial
x}\lambda(x,t)\rho(x,t)dx\nonumber
=-E\left\{\frac{\partial}{\partial
x}\lambda(x(t),t)\right\},\nonumber
\end{eqnarray}
where, in the integration by parts, we have used the natural boundary
condition at
infinity for $\rho(\cdot,t)$. Thus, our $\Lambda^{F,\lambda}$
functionals now have the form
\begin{eqnarray}	\nonumber &&\Lambda^{F,\lambda}(x,v,u') =E\{
F(x(0),0)-F(x(-T),-T)\\&& -
\int_{-T}^{0}\left\{\left[\frac{\partial{F}}{\partial t} +
v(t)\frac{\partial}{\partial x}
F\right](x(t),t)-\lambda(x(t),t)\cdot
u'(t)-\frac{\partial}{\partial
x}\lambda(x(t),t)\right\}dt\}.\nonumber
\end{eqnarray}
The variational analysis now follows the same lines as in \cite{P3}.
Consider now the {\it unconstrained}  problem
$$\min_{x\in {\cal X}_0}\min_{v\in {\cal V}}\max_{u'\in {\cal
V}} (I +
\Lambda ^{F,\lambda})(x,v,u').
$$
For each fixed $x\in {\cal X}_0$, and each $t\in
[-T,0]$, we study the {\it finite-dimensional} problem

\begin{eqnarray}\nonumber
\min_{v\in \R}\max_{u'\in
\R}\left\{\frac{1}{2}m(v-\frac{1}{m}\frac{\partial}{\partial
x}
S_r(x(t),t))^2-
\frac{\hbar^2}{8m}(u'-\frac{\partial}{\partial
x}\log\rho_r(x(t),t))^2\right.\\\left.
-\frac{\partial{F}}{\partial t}(x(t),t) - v \frac{\partial}{\partial
x}
F(x(t),t)+\lambda(x(t),t)
u'+\frac{\partial}{\partial
x}\lambda(x(t),t)\right\}.\nonumber\end{eqnarray}
We get the optimality
conditions
\begin{eqnarray}\nonumber		 v_x^*(t)
=\frac{1}{m}\frac{\partial}{\partial
x}
S_r(x(t),t))+\frac{1}{m}\frac{\partial}{\partial
x} F(x(t),t),
\\u'^*_x(t)=\frac{\partial}{\partial
x}\log\rho_r(x(t),t))+\frac{4m}{\hbar^2}\lambda(x(t),t
). \nonumber
\end{eqnarray}

\begin{remark} {\em If a stochastic
process with the prescribed $v_x^*(t)$ and $u'^*_x(t)$ does exist, then the
first optimality condition implies  that it is a
{\em Markov} process and  the second that $\lambda$
is given by
$$\lambda(x,t)=\frac{\hbar^2}{4m}\frac{\partial}{\partial
x}\log\left(\frac{\rho^*_x}{\rho_
r}\right)(x,t),
$$
where $\rho^*_x(x,t)$ denotes the probability density of $x^*(t)$.}
\end{remark}
Notice that  $v^x$ and $u'^x$ belong to ${\cal V}$.
Consider next the minimization of
\begin{eqnarray}\nonumber (I + \Lambda^{F,\lambda})(x,v_x^*,u'^*_x)=E\left.\{
F(x(0),0)-F(x(-T),-T)\right.\\\nonumber\left.
+\int_{-T}^{0}\left[-\frac{\partial F}{\partial
t}(x(t),t)-\frac{1}{2m}\left(\frac{\partial}{\partial
x} F(x(t),t)\right)^2-\frac{1}{m}\frac{\partial}{\partial
x} S_r(x(t),t))\frac{\partial}{\partial
x} F(x(t),t)
\nonumber\right.\right.\\\left.\left.
\frac{2m}{\hbar^2}\lambda(x(t),t)^2
+\frac{\partial}{\partial
x}\log\rho_r(x(t),t))\cdot\lambda(x(t),t)+
\frac{\partial}{\partial
x}\cdot\lambda(x(t),t)\right]dt\right\}\nonumber\end{eqnarray} on the space
${\cal X}_0$.
We wish to choose $F$ and $\lambda$ such that the functional becomes
constant with
respect to the process $x\in{\cal X}_0$.
Suppose that the pair $(F,\lambda)$
satisfies on $\R\times [-T,0]$ the equation
\begin{equation}\label{1}\frac{\partial F}{\partial
t}+\frac{1}{2m}\left(\frac{\partial}{\partial
x} F\right)^2+\frac{1}{m}\frac{\partial}{\partial
x} S_r\cdot\frac{\partial}{\partial
x}
F -
\frac{2m}{\hbar^2}\lambda^2-\frac{\partial}{\partial
x}\log\rho_r\cdot\lambda-
\frac{\partial}{\partial
x}\cdot\lambda= 0,
\end{equation}
and the boundary condition $F(x,t_0) = -S_r(x,0)$.
With this choice of $(F,\lambda)$,
$(I+\Lambda^{F,\lambda})(x,v_x^*,u'^*_x))\equiv E\{F(x(0),0)\}$
which is constant on ${\cal
X}_0$ (all processes have the same marginal density at $t=0$).
Hence, {\it any} $x$ in ${\cal
X}_0$ solves with $(v^x,u'^x)$, the unconstrained problem.
\section{Solution to the variational problem}

In view of Remark 4.1, we now write
$$\lambda(x,t)=\frac{\hbar^2}{2m}\frac{\partial}{\partial
x} G(x,t),
$$
for some scalar $C^1$ function $G$. Equation (\ref{1}) then becomes
\begin{equation}\label{2}\frac{\partial F}{\partial
t}+\frac{1}{2m}\left(\frac{\partial}{\partial
x} F\right)^2+\frac{1}{m}\frac{\partial}{\partial
x} S_r\cdot\frac{\partial}{\partial
x}
F -
\frac{\hbar^2}{2m}\left[\left(\frac{\partial}{\partial
x} G\right)^2+\frac{\partial^2}{\partial
x^2}
G+\frac{\partial}{\partial
x}\log\rho_r\frac{\partial}{\partial
x} G\right]= 0.
\end{equation}
In the ideal case where $\frac{\partial}{\partial
x} S_r\equiv 0$ and
$\rho_r={\rm const}$, this equation reduces
to one of the Madelung equations. If  $F$ and $G$ may be found
satisfying (\ref{2}) with $F(x,t_0) = -S_r(x,0)$,
then {\it any} $x$ in ${\cal
X}_{\rho_1}$ solves the unconstrained problem.
If we can find one $x^*$ in ${\cal
X}_{\rho_1}$ that also satisfies the constraints
\begin{eqnarray}	\nonumber v_x^*(t) =\frac{1}{m}\frac{\partial}{\partial
x}
S_r(x(t),t))+\frac{1}{m}\frac{\partial}{\partial
x} F(x(t),t) \;{\rm is \;the \;current\;
drift\; of}
\;x,\\ \frac{\partial}{\partial
x}\log\rho^*(x(t),t)=\frac{\partial}{\partial
x}\log\rho_r(x(t),t)+2\frac{\partial}{\partial
x}
G(x,t)\;\forall t,\nonumber
\end{eqnarray}
then it solves the original stochastic differential game. In that
case, we define
\begin{equation}\label{3}S^*(x,t):=S_r(x,t)+F(x,t),
\end{equation}
so that $v_x^*(t) =\frac{1}{m}\frac{\partial}{\partial
x} S^*(x(t),t))$ and we have
\begin{equation}\label{4}
G(x,t)=\frac{1}{2}\log\frac{\rho^*_x}{\rho_r}(x,t).
\end{equation}
Suppose that we
have found such a process $x^*$ with density $\rho^*$. As observed
before, it is a Markov process and so is the
Nelson reference process $x_r$. Then the corresponding Fokker-Planck
(continuity) equations read
\begin{eqnarray}
\frac{\partial \rho^*}{\partial t}+\frac{\partial}{\partial
x}(\frac{1}{m}\frac{\partial}{\partial
x} S^* \rho^*)=0,
\\\frac{\partial \rho_r}{\partial
t}+\frac{\partial}{\partial
x}(\frac{1}{m}\frac{\partial}{\partial
x} S_r \rho_r)=0.
\end{eqnarray}
Using these equations, and relations (\ref{3})-(\ref{4}), we get that
$F$ and $G$ must satisfy
another equation
\begin{eqnarray}\nonumber
\frac{\partial G}{\partial t}=\frac{\partial
(\frac{1}{2}\log\frac{\rho^*_x}{\rho_r})}{\partial t}
=\frac{1}{2}\left[-\frac{1}{\rho*}\frac{\partial}{\partial
x}(\frac{1}{m}\frac{\partial}{\partial
x} S^* \rho^*)
+\frac{1}{\rho_r}\frac{\partial}{\partial
x}(\frac{1}{m}\frac{\partial}{\partial
x} S_r \rho_r)\right]\\\nonumber
=-\frac{1}{2m}\frac{\partial^2}{\partial
x^2} (S^*-S_r)-\frac{1}{2m}\frac{\partial}{\partial
x}
S^*\frac{\partial}{\partial
x}\log\frac{\rho^*_x}{\rho_r}
-\frac{1}{2m}\frac{\partial}{\partial
x} (S^*-S_r)\frac{\partial}{\partial
x}\log\rho_r\\\nonumber
=-\frac{1}{2m}\frac{\partial^2}{\partial
x^2} F-\frac{1}{m}\frac{\partial}{\partial
x} F\frac{\partial}{\partial
x}
G+\frac{1}{m}\frac{\partial}{\partial
x} S_r\frac{\partial}{\partial
x} G
-\frac{1}{2m}\frac{\partial}{\partial
x} F\frac{\partial}{\partial
x}\log\rho_r.
\end{eqnarray}
Thus, $F$ and $G$ also satisfy
\begin{equation}\label{5}
\frac{\partial G}{\partial t}+\frac{1}{m}\frac{\partial}{\partial
x} F\frac{\partial}{\partial
x}
G+\frac{1}{2m}\frac{\partial^2}{\partial
x^2} F
+\frac{1}{2m}\frac{\partial}{\partial
x} F\cdot\frac{\partial}{\partial
x}\log\rho_r-\frac{1}{m}\frac{\partial}{\partial
x}
S_r\frac{\partial}{\partial
x} G=0.
\end{equation}
In the ideal case where $\frac{\partial}{\partial
x} S_r\equiv 0$ and
$\rho_r={\rm const}$, this equation reduces
to the other Madelung equation.
Moreover,
\begin{equation}\label{6}
G(x,0)=\frac{1}{2}\log\frac{\rho_0}{\rho_r}(x,0).
\end{equation}
Define $\theta=\exp (G+\frac{i}{\hbar}F)$. Then, equations
(\ref{2})-(\ref{5}) imply that
$\theta$ satisfies
\begin{equation}\label{7}
\frac{\partial{\theta}}{\partial{t}} +\left(\frac{1}{m}\frac{\partial}{\partial
x} S_r
+\frac{\hbar}{im}\frac{\partial}{\partial
x} R_r\right)
\frac{\partial}{\partial
x} \theta
-\frac{i\hbar}{2m}\frac{\partial^2}{\partial
x^2}\theta=0
\end{equation}
Define now
$\psi_n:=(\rho^*)^{1/2}\exp{\frac{i}{\hbar}S^*}=\psi_r\theta$. It
follows from
(\ref{0}) and (\ref{7}) that $\psi_n$ also satisfies (\ref{0})
and $\psi_n(x,0)=(\rho_0(x))^{1/2}$.
\section{Conclusion}

Hence, we find that
the solution to the stochastic differential game is the Nelson
process associated to another solution of
the same Schr\"{o}dinger equation (\ref{0}), with the
probability density $\rho_0$ at
time $0$ and  $S^*(x,0)=0$ (so that $v(0)=0$).  In other words, the
variational principle shows that the process with density $\rho_0$ at
time $0$ and $v(0)=0$ that replaces the one-slit reference process
is precisely the two-slit process studied by Nelson in \cite{N2,N3}.

As observed by Nelson in
\cite{N3}, the probability density $\rho_0(x)$ can, for all practical purposes,
be replaced by $|\psi_1(x,0)|^2$, where $\psi_1$ was defined in (\ref{2s}).
Thus, the new process, obtained from the variational principle, will
exhibit, after
a time sufficiently large with respect to $a$, the typical
interference pattern in
the probability density.

We have made here no attempt at a physical interpretation of the variational
mechanism. We refer the reader to \cite{G94,G97,CPM}, and references 
therein, for a
discussion of various crucial topics in stochastic mechanics from the physical
standpoint.

\vspace{.5cm}

\section*{Acknowledgment}
 
\noindent
I wish to thank Laura Morato for several useful conversations on the
two-slit experiment.

\end{document}